\documentclass[12pt]{extarticle}
\pdfoutput=1 
\usepackage{slashed}
\usepackage{color,colortbl,verbatim}
\usepackage{latexsym}
\usepackage{amssymb}
\usepackage{amsmath}
\usepackage{graphicx}
\graphicspath{{figs/}}
\usepackage{arydshln}
\usepackage[dvipsnames]{xcolor}
\usepackage{adjustbox}
\usepackage{easybmat}
\usepackage{bbm}
\usepackage{cite}
\usepackage{slashed}
\usepackage{bm}
\usepackage{adjustbox}
\usepackage{booktabs}
\usepackage{multirow} 
\usepackage{rotating}
\usepackage{mathtools}
\usepackage[inline]{enumitem}
\usepackage{lscape}
\usepackage{booktabs}
\usepackage[normalem]{ulem}
\usepackage{hyperref}
\usepackage{bm}

\definecolor{LightCyan}{rgb}{0.88,1,1}

\newcommand{\ii}{\ensuremath{\mathrm{i}}}
\renewcommand{\theequation}{\thesection.\arabic{equation}}
\makeatletter
\@addtoreset{equation}{section}
\g@addto@macro\bfseries{\boldmath}
\newcommand\Label[1]{&\refstepcounter{equation}(\theequation)\ltx@label{#1}&}
\makeatother
\allowdisplaybreaks

\begin{document}
%
\thispagestyle{empty}
\begin{flushright}
\end{flushright}
\vspace{0.8cm}

\begin{center}
{\Large\sc Constraints on anomalous dimensions \\[0.3cm] from the positivity of the S-matrix}
\vspace{0.8cm}

\textbf{
Mikael Chala}\\
\vspace{1.cm}
{\em {Departamento de F\'isica Te\'orica y del Cosmos,
Universidad de Granada, Campus de Fuentenueva, E--18071 Granada, Spain}}
\vspace{0.5cm}
\end{center}
\begin{abstract}
We show that the analyticity and crossing symmetry of the S-matrix, together with the optical theorem, impose restrictions on the renormalisation group evolution of dimension-eight operators in the Standard Model Effective Field Theory. Moreover, in the appropriate basis of operators, the latter manifest as zeros in the anomalous dimension matrix that, to the best of our knowledge, have not been anticipated anywhere else in the literature. Our results can be trivially extended to other effective field theories.
\end{abstract}

\newpage

\tableofcontents

\section{Introduction}
One of the most studied aspects of quantum field theory (QFT) is the evolution of the scale-dependent parameters under renormalisation group (RG) running. For renormalisable QFTs involving only scalars, fermions and gauge bosons, the explicit form of the RG equations (RGEs) is completely known up to two loops~\cite{Machacek:1983tz,Machacek:1983fi,Machacek:1984zw,Jack:1982sr,Luo:2002ti,Lyonnet:2013dna,Lyonnet:2016xiz,Luo:2002iq,Fonseca:2013bua,Schienbein:2018fsw,Sartore:2020gou}.  For QFTs involving operators of dimension larger than four, also known as effective field theories (EFTs), this problem is significantly much more complicated. Besides the larger number of interactions, the reason is the ubiquitous mixing between different parameters $c_i$, described by the anomalous dimensions matrix (ADM) $\gamma$; $d c_i/d\log{\tilde{\mu}} = \gamma_{ij} c_j$ if we stick to operators of the same dimension ($\tilde{\mu}$ stands for the renormalisation scale).

Since the last ten years or so, there has been significant progress towards the renormalisation of EFTs, particularly in the Standard Model (SM) EFT (SMEFT)~\cite{Grzadkowski:2010es,Brivio:2017vri} at one loop and up to dimension eight~\cite{Jenkins:2013wua,Jenkins:2013zja,Alonso:2013hga,Alonso:2014zka,Liao:2016hru,Davidson:2018zuo,Chala:2021juk,Chala:2021pll,AccettulliHuber:2021uoa,DasBakshi:2022mwk,Helset:2022pde,Asteriadis:2022ras,DasBakshi:2023htx}. Software tools that automatise part of this process have been of enormous importance in this respect~\cite{Hahn:1998yk,Hahn:2000kx, Carmona:2021xtq}. Still, the computations entail so many technical and conceptual challenges, that the full renormalisation of arbitrary EFTs is far from complete.

However, there are \textit{aspects} of the RG flow of EFTs that can be understood without necessarily struggling with the explicit calculation of RGEs. One of this aspects is the existence of fixed points in the RG flow, or zeros of the ADM. Approaches based on generalised unitarity~\cite{Bern:1994cg,Bern:1994zx}, together with on-shell amplitude methods~\cite{Cheung:2015aba}, have shown that certain classes of operators do not mix under RG running, despite the fact that, within the Feynman approach to renormalisation, there are diagrams that are separately non-vanishing. These techniques have been successfully applied at one-loop to EFTs of dimensions five and six~\cite{Cheung:2015aba} and eight~\cite{Craig:2019wmo}. See also Refs.~\cite{Elias-Miro:2014eia} and \cite{Helset:2022pde} for works that unveil certain non-trivial zeros in the ADM of the SMEFT, but from the perspectives of Supersymmetry and of EFT geometry, respectively.

In this paper, we want to focus on a related but different aspect of the RG flow, namely on the signs of the anomalous dimensions, which dictate the increase or decrease of the EFT couplings under running. To this aim, we rely on constraints on the forward amplitude of two-to-two processes that ensue from the principles of crossing symmetry, locality and unitarity of the S-matrix~\cite{Adams:2006sv}, focusing on the SMEFT at dimension eight. To the best of our knowledge, none or very little is known about this (clearly important) facet of EFTs.

The article is structured as follows. We introduce the basics of the SMEFT in  section~\ref{sec:eft}. In section~\ref{sec:dispersion_relations}, we work out different dispersion relations relating infrared (IR) and ultraviolet (UV) properties of renormalisable extensions of the SM, and use them to constrain the behaviour of certain beta functions. In section~\ref{sec:anomalous_dimensions} we thoroughly apply these results to unravel a number of relations between different anomalous dimensions in a simplified version of the SMEFT. We extend this procedure to an almost complete version of the SMEFT in section~\ref{sec:full_smeft}, where we also discuss the appearance of new zeros in the ADM, pertaining not to the mixing between different classes of operators, but rather to the mixing between concrete operators of different classes. We conclude in section~\ref{sec:conclusions}. We dedicate appendix~\ref{app:xcheck} to cross-check the validity of \textit{some} of our findings by explicit calculation.

\section{Effective field theory}
\label{sec:eft}
\renewcommand{\arraystretch}{1.5}

\begin{table}[!ht]
 \begin{center}
  \resizebox{\textwidth}{!}{
  \begin{tabular}{cclcl}
   \toprule
   & \textbf{Operator} & \textbf{Notation} & \textbf{Operator} & \textbf{Notation}\\[0.5cm]
   \boldmath{$\phi^8$} & $( \phi^\dagger\phi)^4$ & $\mathcal{O}_{\phi^8}$ & &\\[0.1cm]
   %
   \boldmath{$\phi^6 D^2$} &  $(\phi^{\dag} \phi)^2 (D_{\mu} \phi^{\dag} D^{\mu} \phi)$ & $\mathcal{O}_{\phi^6}^{(1)}$  &  $(\phi^{\dag} \phi) (\phi^{\dag} \sigma^I \phi) (D_{\mu} \phi^{\dag} \sigma^I D^{\mu} \phi)$ & $\mathcal{O}_{\phi^6}^{(2)}$ \\[0.1cm]
   %
   \rowcolor{LightCyan}
   \multirow{2}{*}{\boldmath{$\phi^4 D^4$}} &  $(D_{\mu} \phi^{\dag} D_{\nu} \phi) (D^{\nu} \phi^{\dag} D^{\mu} \phi)$ &  $\mathcal{O}_{\phi^4}^{(1)}$ &  $(D_{\mu} \phi^{\dag} D_{\nu} \phi) (D^{\mu} \phi^{\dag} D^{\nu} \phi)$ & $\mathcal{O}_{\phi^4}^{(2)}$ \\[0.2cm]
   \rowcolor{LightCyan}
   & $(D^{\mu} \phi^{\dag} D_{\mu} \phi) (D^{\nu} \phi^{\dag} D_{\nu} \phi)$ & $\mathcal{O}_{\phi^4}^{(3)}$ & & \\[0.1cm]
   %
   \boldmath{$B^2 \phi^4$} & $ (\phi^\dag \phi)^2 B_{\mu\nu} B^{\mu\nu}$ & $\mathcal{O}_{B^2\phi^4}^{(1)}$ & $(\phi^\dag \phi)^2 \widetilde B_{\mu\nu} B^{\mu\nu}$ & $\mathcal{O}_{B^2\phi^4}^{(2)}$\\[0.1cm]
   \boldmath{$B \phi^4 D^2$} &  $\text{i}(\phi^{\dag} \phi) (D^{\mu} \phi^{\dag} D^{\nu} \phi) B_{\mu\nu}$ & $\mathcal{O}_{B\phi^4D^2}^{(1)}$ & $\text{i}(\phi^{\dag} \phi) (D^{\mu} \phi^{\dag} D^{\nu} \phi) \widetilde{B}_{\mu\nu}$ & $\mathcal{O}_{B\phi^4D^2}^{(2)}$ \\[0.1cm]
   %
   \rowcolor{LightCyan}
   \multirow{2}{*}{\textcolor{gray}{\boldmath{$B^2\phi^2 D^2$}}} & \textcolor{gray}{$(D^\mu\phi^\dagger D^\nu\phi) B_{\mu\rho} B_\nu^\rho$} & \textcolor{gray}{$\mathcal{O}_{B^2\phi^2 D^2}^{(1)}$} & \textcolor{gray}{$(D^\mu\phi^\dagger D^\mu\phi) B_{\nu\rho} B^{\nu\rho}$} & \textcolor{gray}{$\mathcal{O}_{B^2\phi^2 D^2}^{(2)}$} \\[0.2cm]
   \rowcolor{LightCyan}
   & \textcolor{gray}{$(D^\mu\phi^\dagger D^\mu\phi) B_{\nu\rho} \tilde{B}^{\nu\rho}$} & \textcolor{gray}{$\mathcal{O}_{B^2\phi^2 D^2}^{(3)}$} & & \\[0.1cm]
   %
   \rowcolor{LightCyan}
   \textcolor{gray}{\multirow{2}{*}{\boldmath{$B^4$}}} & \textcolor{gray}{$(B_{\mu\nu}B^{\mu\nu})(B_{\rho\sigma} B^{\rho\sigma})$} & \textcolor{gray}{$\mathcal{O}_{B^4}^{(1)}$} & \textcolor{gray}{$(B_{\mu\nu}\tilde{B}^{\mu\nu})(B_{\rho\sigma} \tilde{B}^{\rho\sigma})$} & \textcolor{gray}{$\mathcal{O}_{B^4}^{(2)}$} \\[0.1cm]
   \rowcolor{LightCyan}
   & \textcolor{gray}{$(B_{\mu\nu}B^{\mu\nu})(B_{\rho\sigma} \tilde{B}^{\rho\sigma})$} & \textcolor{gray}{$\mathcal{O}_{B^4}^{(3)}$} & & \\[0.1cm]
   %
   \rowcolor{LightCyan}
   \boldmath{$e^2 \phi^2 D^3$} & $\text{i} (\overline{e}\gamma^\mu D^\nu e)(D_{(\mu} D_{\nu)}\phi^\dagger\phi) + \text{h.c.}$ & $\mathcal{O}_{e^2\phi^2 D^3}^{(1)}$ & $\text{i} (\overline{e}\gamma^\mu D^\nu e)(\phi^\dagger D_{(\mu} D_{\nu)}\phi) + \text{h.c.}$ & $\mathcal{O}_{e^2\phi^2 D^3}^{(2)}$ \\[0.1cm]
   %
   \boldmath{$e^2 \phi^4 D$} & $\text{i} (\overline{e}\gamma^\mu e)(\phi^\dagger D_\mu\phi)(\phi^\dagger\phi)+\text{h.c.}$ & $\mathcal{O}_{e^2\phi^4 D}$ & & \\[0.1cm]
   %
   \rowcolor{LightCyan}
   \textcolor{gray}{\boldmath{$e^2 B^2 D$}} & \textcolor{gray}{$\text{i} (\overline{e}\gamma^\mu D^\nu e) B_{\mu\rho} B_\nu^\rho + \text{h.c.}$} & \textcolor{gray}{$\mathcal{O}_{e^2 B^2 D}$} & & \\[0.1cm]
   %
   \multirow{2}{*}{\boldmath{$e^2 B \phi^2 D$}} & $(\overline{e}\gamma^\nu e) D^\mu (\phi^\dagger\phi) B_{\mu\nu}$ & $\mathcal{O}_{e^2 B\phi^2 D}^{(1)}$ & $(\overline{e}\gamma^\nu e) D^\mu (\phi^\dagger\phi) \tilde{B}_{\mu\nu}$ & $\mathcal{O}_{e^2 B\phi^2 D}^{(2)}$\\[0.1cm]
   & $(\overline{e}\gamma^\nu e) (\phi^\dagger \overleftrightarrow{D}^\mu\phi) B_{\mu\nu}$ & $\mathcal{O}_{e^2 B\phi^2 D}^{(3)}$ & $(\overline{e}\gamma^\nu e) (\phi^\dagger \overleftrightarrow{D}^\mu\phi) \tilde{B}_{\mu\nu}$ & $\mathcal{O}_{e^2 B\phi^2 D}^{(4)}$ \\[0.1cm]
   %
   \boldmath{$e^4 \phi^2$} & $(\overline{e}\gamma^\mu e)(\overline{e}\gamma_\mu e) (\phi^\dagger\phi)$ & $\mathcal{O}_{e^4\phi^2}$ & &\\[0.1cm]
   %
   %
   \rowcolor{LightCyan}
   \boldmath{$e^4 D^2$} & $D^\nu (\overline{e}\gamma^\mu e)D_\nu (\overline{e}\gamma_\mu e)$ & $\mathcal{O}_{e^4 D^2}$ & & \\[0.1cm]
   \bottomrule
  \end{tabular}
  }
 \end{center}
 \caption{\it Dimension-8 operators in the rSMEFT. Operators in \textcolor{gray}{gray} arise only at loop level in weakly-coupled UV completions of the SMEFT. \fcolorbox{white}{LightCyan}{Shaded} operators contribute to two-to-two amplitudes, being subject to positivity bounds. Note that operators of the type $e^4 B$ are absent for only one family of leptons.}\label{tab:rSMEFT}
\end{table}
We use the following notation for the SM fields: $e$, $u$ and $d$ denote the right-handed leptons, up and down quarks, respectively, and $l$ and $q$ represent the left-handed counterparts; $B$, $W$ and $G$ refer to the electroweak gauge bosons and the gluon, respectively, and $g_1$, $g_2$ and $g_3$ are the corresponding gauge couplings; $\phi=(\varphi_1+\ii\varphi_2,\varphi_3+\ii \varphi_4)^T$ stands for the Higgs doublet. Thus, the SM Lagrangian reads:
\begin{align}
 \mathcal{L}_\text{SM} &= -\frac{1}{4}G_{\mu\nu}^A G^{A\,\mu\nu}-\frac{1}{4}W^I_{\mu\nu}W^{I\,\mu\nu}-\frac{1}{4}B_{\mu\nu}B^{\mu\nu}\\\nonumber
 &+\overline{q}\ii\slashed{D}q+\overline{l}\ii\slashed{D} l+\overline{u}\ii\slashed{D}u+\overline{d}\ii\slashed{D}d+\overline{e}\ii\slashed{D}e\\\nonumber
 &+ (D_\mu\phi)^\dagger (D^\mu\phi)+\mu_\phi^2|\phi|^2 -\lambda_\phi |\phi|^4- (\overline{q} \tilde{\phi} Y_u u + \overline{q}\phi Y_d d+\overline{l}\phi Y_e e + \text{h.c.})\,.
\end{align}
The Yukawa couplings $Y_u$, $Y_d$ and $Y_e$ are matrices in flavour space, and $\tilde{\phi}=\ii\sigma_2\phi^*$ with $\sigma_I$ being the Pauli matrices; $I=1,2,3$.

In the absence of lepton-number violation, the SMEFT Lagrangian reads:
\begin{equation}
 \mathcal{L}_\text{SMEFT} = \mathcal{L}_\text{SM} + \frac{1}{\Lambda^2} \sum_i c_i^{(6)}\mathcal{O}_i^{(6)} + \frac{1}{\Lambda^4}\sum_j c_j^{(8)}\mathcal{O}_j^{(8)}+ \cdots\,,
\end{equation}
where $\Lambda\gg 100$ GeV represents the cut-off below which the SMEFT is no longer a valid theory, and
the ellipses encode higher-dimensional operators.
The first sum runs over a basis of dimension-six interactions~\cite{Grzadkowski:2010es}, while the second does it over the dimension-eight counterpart~\cite{Murphy:2020rsh,Li:2020gnx}. In this work, we are mostly interested in the dimension-eight Wilson coefficients. The dependence of these on the energy scale $\tilde{\mu}$ is governed by the corresponding beta functions, which at one loop read:
\begin{equation}\label{eq:running}
 \dot{c}_i^{(8)} = 16\pi^2 \tilde{\mu} \frac{d c_i^{(8)}}{d\tilde{\mu}} = \gamma_{ij} c_j^{(8)} + \gamma_{ijk}^\prime c_j^{(6)} c_k^{(6)}\,.
\end{equation}
(We also use the common notation $\beta_i$ for $\dot{c}_i$.)
A notable part of the ADM $\gamma$ has been already computed explicitly in Refs.~\cite{AccettulliHuber:2021uoa,DasBakshi:2022mwk,Helset:2022pde}, but most is still missing. Likewise for $\gamma'$; see Refs.~\cite{Chala:2021pll,Helset:2022pde,Asteriadis:2022ras}. In this work, we want rather to unveil certain correlations (some known, some others not previously anticipated) between different entries in $\gamma$ without relying on explicit calculations.

Hereafter, and until section~\ref{sec:full_smeft}, we focus on a simplified version of the SM and of its effective extension, that we refer to as reduced SMEFT (rSMEFT), in which the only degrees of freedom are one family of $e$ and the $B$, and the only non-vanishing SM coupling is $g_1$. This EFT is simple enough to be described with a single and digestible table of operators even at dimension eight (see Tab.~\ref{tab:rSMEFT}), but sufficiently rich to capture all the nuances of the interplay between positivity and running in the SMEFT.

\section{Dispersion relations and running}
\label{sec:dispersion_relations}
Let us consider a renormalisable completion of the rSMEFT, with only new particles of mass $M\gtrsim \Lambda$. We use $\psi$ and $\Psi$ when referring collectively to light and heavy fields, respectively.
We are interested in the Wilson coefficients of \textit{four-field} rSMEFT operators generated by integrating out $\Psi$ at tree level. These are independent of $g_1$, so they can be computed in the limit $g_1=0$, which we assume until otherwise stated.

\begin{figure}[t]
 \hspace{0.5cm}
 \includegraphics[width=0.49\columnwidth]{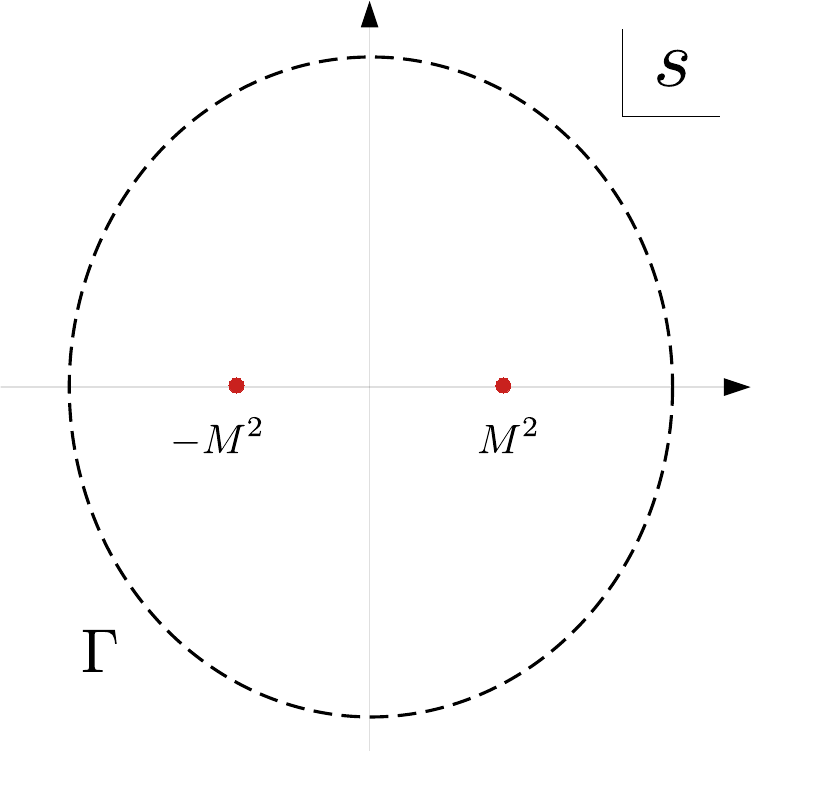}
 \includegraphics[width=0.49\columnwidth]{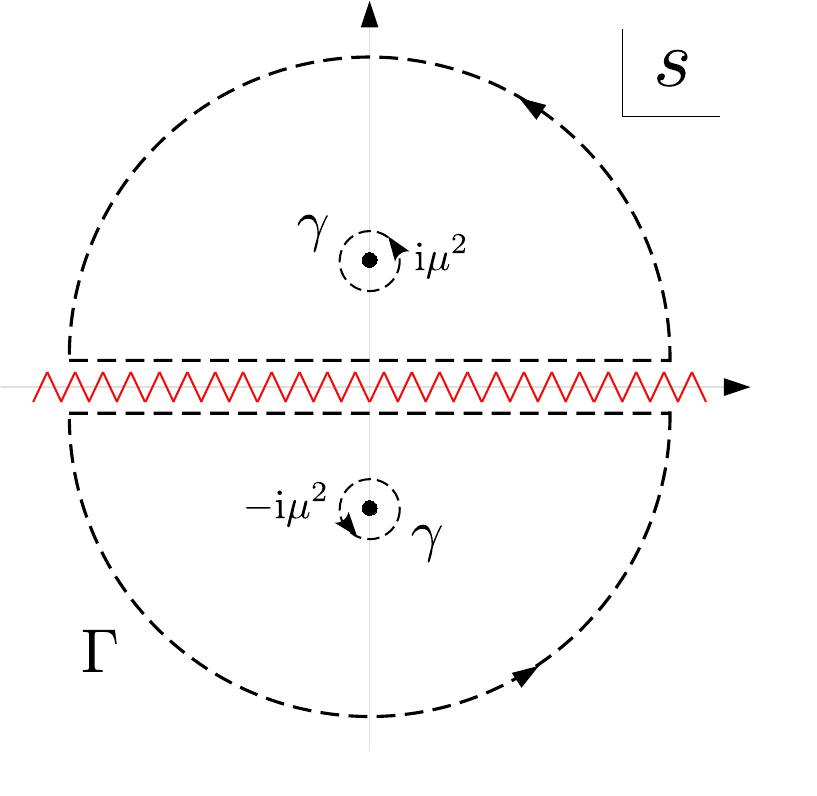}
 \caption{\it Structure of the singularities of a two-to-two amplitude in the forward limit in the plane of the complex Mandelstam variable $s$ at tree level (left) and at one loop (right).}\label{fig:contours}
\end{figure}

Following Ref.~\cite{Adams:2006sv}, we consider a crossing-symmetric elastic process $\psi_i\psi_j\to \psi_i\psi_j$ at tree level with amplitude $\mathcal{A}(s,t) = \mathcal{A}(u,t)$, where $s,t$ and $u$ are the Mandelstam variables; $s+t+u=0$ because all light particles are massless. In the forward limit $t\to 0$, $u\to -s$, and therefore $\mathcal{A}(s)\equiv\mathcal{A}(s,t=0)$ is symmetric in $s$; $\mathcal{A}(s)=\mathcal{A}(-s)$. Promoting $s$ to a complex variable, the analytic structure of $\mathcal{A}(s)$ consists simply of two poles at $s=\pm M^2$. Poles at $s=0$, ensuing for example from the exchange of a gauge boson in Higgs scattering, are absent in the regime $g_1=0$. For the same reason, the forward limit is well defined. Using Cauchy's theorem, we can write:
\begin{equation}\label{eq:dispersion_tree}
 \mathcal{I} \equiv \int_\Gamma \frac{\mathcal{A}(s)}{s^3} = \text{Res}\,\frac{\mathcal{A}}{s^3}(s=0) + 2 \,\frac{1}{M^3}\text{Res}\,\mathcal{A}(s=M^2)\,,
\end{equation}
where $\Gamma$ is a circular path extending to arbitrarily large values of $|s|$ (see Fig.~\ref{fig:contours}), and the factor of $2$ comes from the symmetry of the amplitude under change of sign in $s$. 

By virtue of to the Froissart's bound~\cite{Froissart:1961ux}, $\mathcal{A}(s)/s^3\to 0$ at infinity, so $\mathcal{I} =0$. Thus, Eq.~\eqref{eq:dispersion_tree} relates an IR quantity with an UV observable. The latter, given by the residue at $s=M^2$, can be shown to be negative. Indeed, by definition, we have that in the limit $s\to M^2$:
\begin{equation}
 \mathcal{A}(s) \to \lim_{\epsilon\to 0}\frac{1}{s-M^2+i\epsilon} \text{Res}\,\mathcal{A}(s=M^2)\,,
\end{equation}
which implies that:
\begin{align}
 \text{Im}\mathcal{A}(s)&\to \lim_{\epsilon\to 0}\frac{-\epsilon}{(s-M^2)^2+\epsilon^2}\text{Res}\,\mathcal{A}(s=M^2)\\\nonumber
 &=-\pi \delta(s-M^2)\text{Res}\,\mathcal{A}(s=M^2)\,.
\end{align}
By the optical theorem (the completion of the rSMEFT is assumed renormalisable and therefore its S-matrix is unitary), the imaginary part of the forward amplitude is positive. Hence, integrating the equation above over a small (positive) interval around $s=M^2$ we obtain that $\text{Res}\,\mathcal{A}(s=M^2) < 0$. Moreover, in the vicinity of $s=0$ the amplitude can be computed within the EFT:
\begin{equation}
 \mathcal{A}(s) = a_0 + a_2 s^2 + a_4 s^4+\cdots\,;
\end{equation}
the residue of $\mathcal{A}(s)/s^3$ at $s=0$ being simply given by $a_2$. Altogether, Eq.~\eqref{eq:dispersion_tree} implies that:
\begin{equation}
 a_2 \geq 0\,.
\end{equation}
For obvious reasons, inequalities of this sort are called \text{positivity bounds}. Applied to particular processes within the rSMEFT, this leads to the following constraints.

For $\varphi_i\varphi_j\to\varphi_i\varphi_j$, we get:
\begin{equation}\label{eq:pos1}
 c_{\phi^4}^{(2)}\geq 0\,, \quad c_{\phi^4}^{(1)}+c_{\phi^4}^{(2)}\geq 0\,,\quad c_{\phi^4}^{(1)}+c_{\phi^4}^{(2)}+c_{\phi^4}^{(3)}\geq 0\,.
\end{equation}

For $\varphi_i B\to\varphi_i B$, we get:
\begin{equation}\label{eq:pos2}
 -c_{B^2\phi^2 D^2}^{(1)} \geq 0\,.
\end{equation}

For $BB\to BB$, we get:
\begin{equation}\label{eq:pos3}
 c_{B^4}^{(1)}\geq 0\,,\quad c_{B^4}^{(2)}\geq 0\,.
\end{equation}

For $e \varphi_i\to e \varphi_i$, we get:
\begin{equation}\label{eq:pos4}
 -c_{e^2\phi^2 D^3}^{(1)}-c_{e^2\phi^2 D^3}^{(2)}\geq 0\,.
\end{equation}

For $e B\to e B$, we get:
\begin{equation}\label{eq:pos5}
 -c_{e^2 B^2 D}\geq 0\,.
\end{equation}

Finally, for $ee\to ee$, we obtain:
\begin{equation}\label{eq:pos6}
 -c_{e^4 D^2}\geq 0\,.
\end{equation}
See also Refs.~\cite{Remmen:2019cyz,Bi:2019phv,Remmen:2020vts,Gu:2020ldn}.

If we restrict our computations to the Higgs unbroken phase, dimension-six operators do not contribute to $a_2$, not even in pairs. This is because, in our basis, all interactions involve at least four fields, and hence pairs of these can only contribute to six-point amplitudes or above.

Let us now assume that, in a particular completion on the rSMEFT, the amplitude considered before vanishes exactly at tree level, but not necessarily at one loop.  The singularity structure of the later consists then of a branch cut extending across the whole $\text{Re}(s)$ axis originated in loops of the massless particles; see Fig.~\ref{fig:contours}. In this case, dispersion relations as that of Eq.~\eqref{eq:dispersion_tree} based on integration contours that cross the $\text{Re}(s)$ axis, can no longer be obtained~\footnote{In principle, we can deform the IR theory by introducing an explicit mass $m$ for the light fields. This way, the branch cut splits into two connected components with branch points starting at $s \sim \pm m^2$. Accordingly, we can consider an integration path that crosses the $\text{Re}(s)$ axis in the vicinity of $s=0$ where the amplitude becomes analytic; see Ref.~\cite{Adams:2006sv}. New dispersion relations can be written from here and, basing again on unitarity, positivity bounds can be obtained. However, translating these bounds to the original massless theory, namely taking the limit $m\to 0$, can be tricky~\cite{Arkani-Hamed:2020blm,Chala:2021wpj}; particularly if they are dominated by the otherwise absent longitudinal degrees of freedom of gauge bosons~\cite{Bellazzini:2016xrt}.

Alternatively, one could consider dispersion relations for subtracted amplitudes in which the low-energy singularities are removed~\cite{Zhang:2021eeo,Li:2022aby}.}. Following Ref.~\cite{Herrero-Valea:2020wxz}, we instead consider the integration paths depicted in the same figure.


We start defining:
\begin{equation}
 \Sigma(\mu) \equiv \frac{1}{2\pi \ii}\int_\gamma \frac{\mathcal{A}(s) s^3}{(s^2+\mu^4)^3} = \frac{1}{2\pi \ii}\int_\Gamma \frac{\mathcal{A}(s) s^3}{(s^2+\mu^4)^3}\,,
\end{equation}
where we have used the analyticity of $\mathcal{A}(s)$ to deform the contour of integration from $\gamma$ to $\Gamma$ in the second equality, again relating an IR quantity with an UV observable. Advocating once more the Froissart's bound, the right-hand side of the equation can be computed explicitly, giving:
\begin{align}
 \Sigma(\mu) &= \frac{1}{\pi\ii}\int_{0}^\infty \frac{s^3}{(s^2+\mu^4)^3}\lim_{\epsilon\to 0}\left[\mathcal{A}(s+\ii\epsilon) -\mathcal{A}(s-\ii\epsilon)\right]\\\nonumber
 &= \frac{1}{\pi\ii}\int_0^\infty \frac{s^3}{(s^2+\mu^4)^3} \lim_{\epsilon\to 0}\left[\mathcal{A}(s+\ii\epsilon)-\mathcal{A}(s+\ii\epsilon)^*\right] = \frac{2}{\pi}\int_0^\infty \frac{\text{Im}\mathcal{A}(s) s^3}{(s^2+\mu^4)^3}\geq 0\,.
\end{align}
In the second equality, we have relied on the Schwarz's reflection principle $\mathcal{A}(s)^*=\mathcal{A}(s^*)$, while in the last one we have used again the optical theorem.

So, $\Sigma(\mu)$ is positive, and from its very definition it can be computed within the EFT provided $\mu\ll\Lambda$. For an amplitude that vanishes at tree level, for any $s$ in the neighborhood of $\pm\ii \mu^2$, we have:
\begin{equation}\label{eq:pos_one_loop}
 \mathcal{A}(s) \sim (\beta_0+\beta_2 s^2 + \beta_4 s^4+\cdots )\log{\frac{s}{\Lambda^2}}\,,
\end{equation}
with $\beta_0$, $\beta_2$, etc. being respectively the beta functions of the dimension-four, dimension-eight, etc. operators in the EFT (not present at tree level in the UV) that contribute to the amplitude at tree level. (In the case of $\beta_0$, this represents a slight abuse of notation, as it can consist of the amplitude obtained by gluing two three-point vertices together.) $\beta_2$ can be simply read off from Eqs.~\eqref{eq:pos1}--\eqref{eq:pos6}; for example, for $e\varphi_i \to e\varphi_i$ in the forward limit, $\beta_2\sim -(\dot{c}_{e^2\phi^2 D^3}^{(1)}+\dot{c}_{e^2\phi^2 D^3}^{(2)})$.

From Eq.~\eqref{eq:pos_one_loop}, we can compute $\Sigma(\mu)$ explicitly by using Cauchy's theorem:
\begin{align}
 \Sigma(\mu) &= \text{Res} \frac{\mathcal{A}(s) s^3}{(s^2+\mu^4)^3}(s=\ii\mu^2) + \text{Res} \frac{\mathcal{A}(s) s^3}{(s^2+\mu^4)^3}(s=-\ii\mu^2) \\\nonumber
 &=\frac{1}{4\mu^4}\left[-\beta_0 + (3+4\log{\frac{\mu^2}{\Lambda^2}})\beta_2\mu^4 - 5 \beta_4 \mu^8 +\cdots \right] +\mathcal{O}( \mu^4\log{\frac{\mu^2}{\Lambda^2}})\,.
\end{align}
%
All terms $\beta_{i\geq 2}$ scale equally with $g_1$, so
there exists a value $\mu'$, independent of $g_1$, below which the $\beta_2$ term in the expression above is larger (in absolute value) than all terms with higher powers of $\mu$. For fixed $\mu$, there exists moreover $g_1'$ for which $\beta_0< \beta_2\,\mu'^4$ for all $g_1\leq g_1'$. This is so because the contributions of the renormalisable operators involve necessarily higher powers of $g_1$~\footnote{For example, the one-loop amplitude for $\varphi_i B\to\varphi_i B$ in the reduced SM scales as $dg_1^2/d\log\mu\sim \mathcal{O}(g_1^4)$, whereas in the EFT we find contributions of the sort $\mathcal{O}(g_1^2\, c_{\phi^4})$. The only exception occurs when the relevant coupling $\lambda_\phi$ is also present in the tree-level EFT, so that $\beta_0$ involves $g_1^2$ corrections too. However, these can be removed, without any further effect, upon fine-tuning the quartic term in the renormalisable Lagrangian.}. Thus, in the double limit $\mu,g_1\to 0$, we have that:
\begin{equation}\label{eq:positive_beta}
 \Sigma(\mu) \approx \log{(\frac{\mu^2}{\Lambda^2})}\beta_2\geq 0\Longrightarrow \beta_2\leq 0\,.
\end{equation}
%


Note that $\beta_2$ involves both the anomalous dimensions $\gamma$ and $\gamma'$, stemming from loops of dimension-eight operators and from loops of pairs of dimension-six ones, respectively; see Eq.~\eqref{eq:running}. The first depends on $g_1$, while the second is $g_1$-independent. Therefore, within the regime $g_1\leq g_1'$ of validity of Eq.~\eqref{eq:positive_beta}, $\gamma_{ij}c_j^{(8)}\leq 0$ can be only implied on a robust basis provided $\gamma'$ vanishes. Luckily enough, this holds in a wide range of cases. For example, in the space of tree-level UV completions of \textit{only} four-Higgs operators (namely $\phi^4 D^4$ but also dimension-six $\phi^4 D^2$ terms), the $\gamma^\prime$ for $B^2\phi^2 D^2$ operators vanishes, because loops with pairs of four-Higgs operators can not contain less than four Higgs external legs. On the contrary, in the space of tree-level UV completions of only two-lepton-two-Higgs interactions, the $\gamma^\prime$ of $\phi^4 D^4$ does not necessarily vanish, because pairs of dimension-six $e^2\phi^2 D$ operators, which are in general present together with $e^2\phi^2 D^3$ terms, can form loops with only four external Higgses. As a general rule, $\gamma^\prime$ can be neglected in the renormalisation of $\psi_i^2\psi_j^2$ operators by $\psi_k^2\psi_l^2$ if either $i\neq k,l$ or $j\neq k,l$. (Remember that $\psi$ stands for any field in the EFT, either fermionic or bosonic.) It can be also ignored in the renormalisation of four-field operators by higher-point interactions.

Now, by simple power counting, one can deduce that, for $\gamma^\prime=0$, Eq.~\eqref{eq:positive_beta} must hold not only for $g_1\leq g_1^\prime$ but also for \textit{all} values of $g_1$. Moreover, here we make the assumption~\footnote{This assumption is supported by substantial evidence in the literature~\cite{Chala:2021wpj,Li:2022rag,Li:2022tcz}. Let us focus, for example, on $\phi^4 D^4$ operators. The extension of the SM with a scalar $\mathcal{S}\sim (1,1)_0$ and two vectors $\mathcal{B}\sim (1,1)_0$ and $\mathcal{B}_1\sim (1,1)_1$ gives $c_{\phi^4}^{(1)}=\delta^2-\gamma^2$, $c_{\phi^4}^{(2)}=\gamma^2$ and $c_{\phi^4}^{(3)}=\alpha^2-\delta^2$, from where: $$c_{\phi^4}^{(2)}=\gamma^2\,,\quad c_{\phi^4}^{(1)}+c_{\phi^4}^{(2)}=\delta^2\,,\quad c_{\phi^4}^{(1)}+c_{\phi^4}^{(2)}+c_{\phi^4}^{(3)}=\alpha^2\,;$$ each of which is obviously arbitrarily non-negative and the three of them are uncorrelated. 

As a matter of fact,  we do not know of constraints stronger than those in Eqs.~\eqref{eq:pos1}--\eqref{eq:pos5} in the space of dimension-eight operators.
} that the positivity bounds can be saturated in the UV, meaning that, for any combination of Wilson coefficients fulfilling Eqs.~\eqref{eq:pos1}--\eqref{eq:pos5}, there exists at least one UV completion of the rSMEFT that leads to this combination in the IR. This implies that Eq.~\eqref{eq:positive_beta} holds for arbitrary values of the tree-level Wilson coefficients satisfying Eqs.~\eqref{eq:pos1}--\eqref{eq:pos6}. We exploit this aspect of positivity, together with the vanishing $\gamma^\prime$ (when possible) in the next section.

\section{Anomalous dimension matrix}
\label{sec:anomalous_dimensions}

Let us focus on the $\gamma$ term appearing in Eq.~\eqref{eq:running} within the rSMEFT. Without further knowledge, and neglecting CP violation for simplicity of the exposition, this is a $20\times 15$ matrix whose entries are completely arbitrary polynomials on $g_1$. (Columns involving operators that only arise at loop level in weakly-coupled UV completions of the rSMEFT, shown in gray in Tab.~\ref{tab:rSMEFT}, are neglected as the running triggered by these interactions is formally a two-loop effect.) In what follows, though, we use the results derived in section~\ref{sec:dispersion_relations} to unravel a number of relations between different entries in the $10\times 15$ sub-matrix involving only rows of operators subject to positivity constraints; see Eqs.~\eqref{eq:pos1}--\eqref{eq:pos5}. This matrix is shown in Tab.~\ref{tab:rges_rSMEFT}. For a clearer reading we do not depict rows and columns that are trivially known to vanish completely. This is the case, for example, of the rows for $B^4$. Indeed, loops involving tree-level operators can not contain \textit{only} $B$s, as they involve at least one external $e$ or $\phi$. For instance, $\phi^4 D^4$ terms can form loops upon closing two Higgses, but other two remain external; likewise two Higgses can be closed in $e^2 B\phi^2 D$ interactions, but then two leptons stay. (Note also that it is irrelevant whether the external $e$ or $\phi$ are on-shell or off-shell, as field redefinitions within the rSMEFT conserve the number of these particles.)

\begin{figure}[t]
 \begin{center}
  \includegraphics[width=0.25\columnwidth]{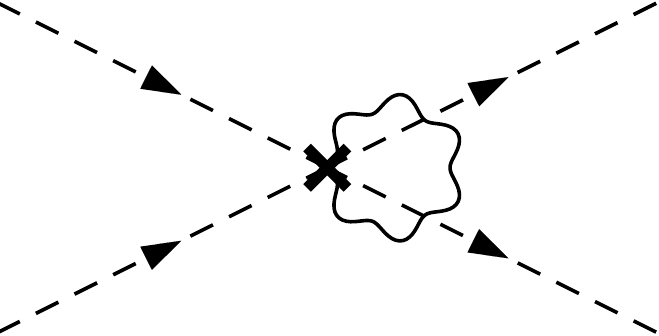}
  \hspace{1cm}
  \includegraphics[width=0.25\columnwidth]{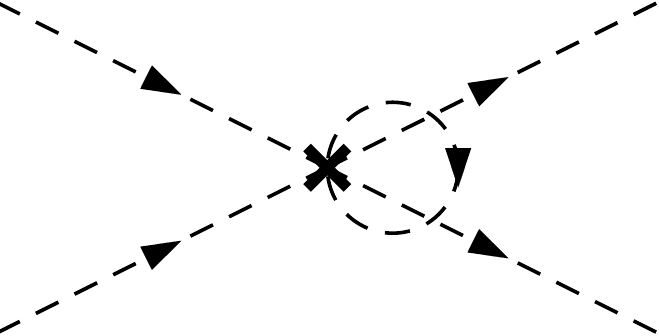}
  \hspace{1cm}
  \includegraphics[width=0.25\columnwidth]{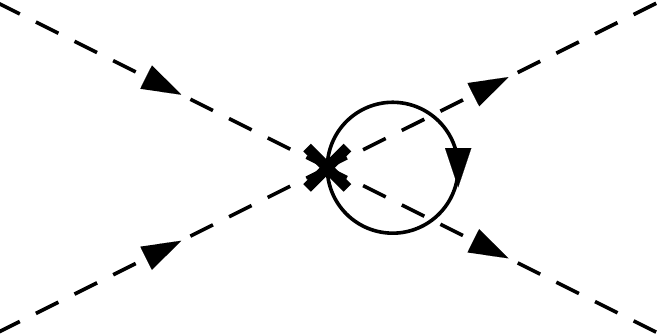}
  \includegraphics[width=0.25\columnwidth]{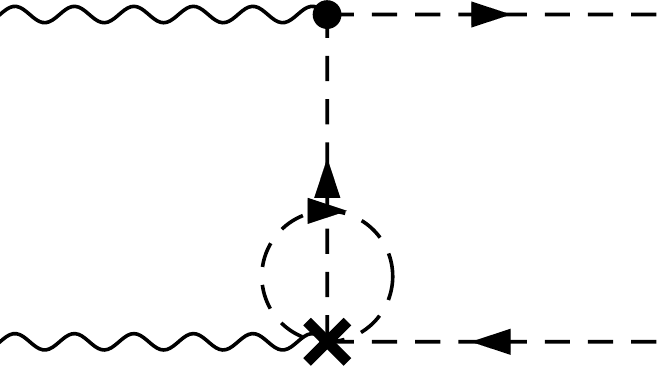}
  \hspace{1cm}
  \includegraphics[width=0.2\columnwidth]{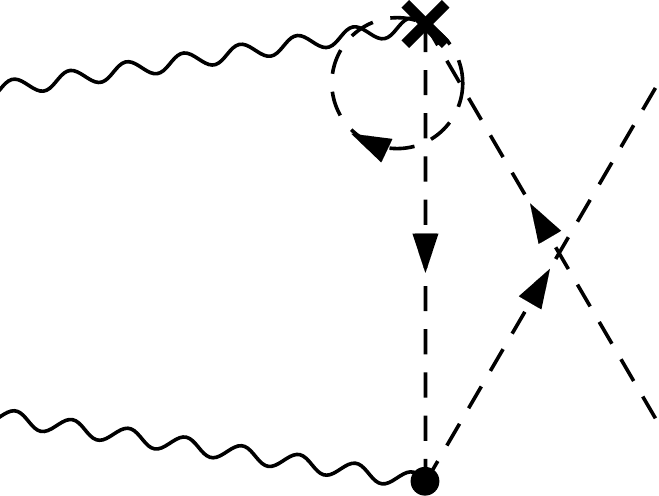}
  \hspace{1cm}
  \includegraphics[width=0.25\columnwidth]{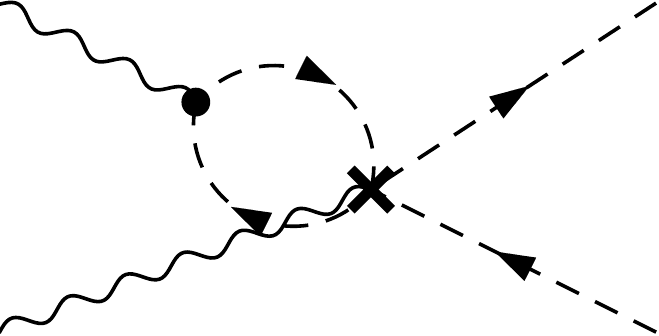}
 \end{center}
 \caption{\it Top) Example diagrams for the renormalisation of $\phi^4 D^4$ by six-field interactions; they all vanish due to the massless bubbles.  Bottom) Example diagrams for the renormalisation of $B^2\phi^2 D^2$ by $B\phi^4 D^2$; the first two vanish because of the massless bubbles, but the third one is non-zero. }\label{fig:diagrams}
\end{figure}

It is also the case of the columns for $\phi^8$, $\phi^6D^2$, $B^2\phi^4$, $e^2\phi^4$ or $e^4\phi^2$. In the first case, upon closing two Higgses to form a loop, six of them remain, while the renormalised operators at hand (those in rows, subject to positivity constraints) contain all four fields only. In the other four cases, loops with only four external fields can be constructed, but they involve bubbles of massless fields, which vanish in the dimensional regularisation scheme that we assume here; see the top panel of Fig.~\ref{fig:diagrams}.

The matrix in Tab.~\ref{tab:rges_rSMEFT} involves still a number of trivial zeros (which do not extend to complete rows or columns though); they are shown as non-shaded. For example, the entry pertaining to the row $c_{\phi^4}^{(1)}$ and column $c_{e^2 B\phi^2 D}^{(1)}$ vanishes because, irrespective of which two legs are closed to form a loop in $e^2 B\phi^2 D$, the resulting diagram contains at least an $e$ or a $B$. (The external $B$s, if off-shell, can be removed at the cost of introducing too many Higgses.)
Other entries, depicted with shaded zeros in Tab.~\ref{tab:rges_rSMEFT}, are \textit{a priori} non-zero, as there exist non-vanishing Feynman diagrams associated to them. For example, the bottom panel of Fig.~\ref{fig:diagrams} shows diagrams for the renormalisation of $B^2\phi^2 D^2$ by $B\phi^4 D^2$. The fact that they also vanish is the first conclusion that can be drawn on the basis of the results obtained in the previous section.

Indeed, from the previous section we know that $\dot{c}_{B^2\phi^2 D^2}^{(1)}$ has definite sign. On the contrary,
$c_{B\phi^4 D^2}^{(1)}$ is not restricted by positivity. Therefore the only option for the renormalisation of $c_{B^2\phi^2 D^2}^{(1)}$ by $c_{B\phi^4 D^2}^{(1)}$ is that $\gamma_{c_{B^2\phi^2 D^2}^{(1)},\, c_{B\phi^4 D^2}^{(1)}}=0$. The rest of the non-trivial zeros on Tab.~\ref{tab:rges_rSMEFT} can be explained on the same footing.

\renewcommand{\arraystretch}{1.5}

\begin{table}[t]
 \begin{center}
  \resizebox{\textwidth}{!}{
  \begin{tabular}{cccccccccc}
   \toprule
   & \textbf{$c_{\phi^4}^{(1)}$} & \textbf{$c_{\phi^4}^{(2)}$} & \textbf{$c_{\phi^4}^{(3)}$} & \textbf{$c_{B\phi^4 D^2}^{(1)}$} & \textbf{$c_{e^2\phi^2 D^3}^{(1)}$} & \textbf{$c_{e^2\phi^2 D^3}^{(2)}$} & \textbf{$c_{e^2 B \phi^2 D}^{(1)}$} & \textbf{$c_{e^2 B\phi^2 D}^{(3)}$}  & \textbf{$c_{e^4 D^2}$} \\[0.5cm]
   \boldmath{$c_{\phi^4}^{(1)}$} & \textcolor{gray}{$\times$} & \textcolor{gray}{$\times$} & \textcolor{gray}{$\times$} & \fcolorbox{black}{LightCyan}{$\,0\,$} & \textcolor{gray}{$\times$} & \textcolor{gray}{$\times$} & 0 & 0 &  0 \\[0.2cm]
   \boldmath{$c_{\phi^4}^{(2)}$} & \textcolor{gray}{$\times$} & \textcolor{gray}{$\times$} & \textcolor{gray}{$\times$} & \fcolorbox{black}{LightCyan}{$\,0\,$} & \textcolor{gray}{$\times$} & \textcolor{gray}{$\times$} & 0 & 0 &  0\\[0.2cm]
   \boldmath{$c_{\phi^4}^{(3)}$} & \textcolor{gray}{$\times$} & \textcolor{gray}{$\times$} & \textcolor{gray}{$\times$} & \fcolorbox{black}{LightCyan}{$\,0\,$} & \textcolor{gray}{$\times$} & \textcolor{gray}{$\times$} & 0 & 0 &  0\\[0.2cm]
   \boldmath{$c_{B^2\phi^2 D^2}^{(1)}$} & \textcolor{blue}{$\bm{\alpha+\beta}$} & \textcolor{blue}{$\bm{\alpha+\beta+\gamma}$} & \textcolor{blue}{$\bm{\alpha}$} & \fcolorbox{black}{LightCyan}{$\,0\,$} & \textcolor{blue}{$\bm{-\delta}$} & \textcolor{blue}{$\bm{-\delta}$} & \fcolorbox{black}{LightCyan}{$\,0\,$} & \fcolorbox{black}{LightCyan}{$\,0\,$} & 0  \\[0.2cm]
   \boldmath{$c_{e^2\phi^2 D^3}^{(1)}$} & \textcolor{blue}{$\bm{\epsilon+\zeta-a$}}  & \textcolor{blue}{$\bm{\epsilon+\zeta+\eta-b}$} & \textcolor{blue}{$\bm{\epsilon-c$}} & 0 & \textcolor{gray}{$\times$} & \textcolor{gray}{$\times$} & \fcolorbox{black}{LightCyan}{$\,0\,$}& \fcolorbox{black}{LightCyan}{$\,0\,$} &  \textcolor{blue}{$\bm{-(\theta+d)}$}\\[0.2cm]
   \boldmath{$c_{e^2\phi^2 D^3}^{(2)}$} & \textcolor{blue}{$\bm{a}$} & \textcolor{blue}{$\bm{b}$} & \textcolor{blue}{$\bm{c}$} & 0 & \textcolor{gray}{$\times$} & \textcolor{gray}{$\times$} & \fcolorbox{black}{LightCyan}{$\,0\,$} & \fcolorbox{black}{LightCyan}{$\,0\,$} &  \textcolor{blue}{$\bm{d}$}\\[0.2cm]
   \boldmath{$c_{e^2 B^2 D}$} & 0 & 0 & 0 & 0 & \textcolor{blue}{$\bm{-\iota}$} & \textcolor{blue}{$\bm{-\iota}$} & \fcolorbox{black}{LightCyan}{$\,0\,$} & \fcolorbox{black}{LightCyan}{$\,0\,$} &  \textcolor{blue}{$\bm{-\kappa}$}\\[0.2cm]
   \boldmath{$c_{e^4 D^2}$} & 0 & 0 & 0 & 0 & \textcolor{gray}{$\times$} & \textcolor{gray}{$\times$} &  0 &  0  & \textcolor{gray}{$\times$}\\[0.2cm]
   \bottomrule
  \end{tabular}
  }
 \end{center}
 \caption{\it Sub-space of the rSMEFT ADM that is constrained by positivity. The Greek letters represent non negative numbers: $\alpha,\beta, ...,\lambda\geq 0$. The Latin letters represent arbitrary numbers. The non-shaded zeros are trivial, while the crosses are unbounded entries.}\label{tab:rges_rSMEFT}
\end{table}

Let us now concentrate on the correlations between different elements appearing on the anomalous dimension matrix. First, we focus on the renormalisation of $B^2\phi^2 D^2$ by $\phi^4 D^4$. According to Eqs.~\eqref{eq:pos2} and \eqref{eq:positive_beta}, we have that $\dot{c}_{B^2\phi^2 D^2}^{(1)}\geq 0$. Following the discussion in the previous section, this inequality must hold for all values of $c_{\phi^4}^{(1)}, c_{\phi^4}^{(2)}, c_{\phi^4}^{(3)}$ compatible with Eq.~\eqref{eq:pos1}. Therefore:
\begin{align}\label{eq:example}
 \dot{c}_{B^2\phi^2 D^2}^{(1)} &= \alpha (c_{\phi^4}^{(1)} +c_{\phi^4}^{(2)}+c_{\phi^4}^{(3)}) + \beta (c_{\phi^4}^{(1)}+c_{\phi^4}^{(2)}) + \gamma c_{\phi^4}^{(2)}+\cdots\\\nonumber
 &=(\alpha+\beta)c_{\phi^4}^{(1)} + (\alpha+\beta+\gamma)c_{\phi^4}^{(2)} + \alpha c_{\phi^4}^{(3)}+\cdots\,,
\end{align}
with $\alpha,\beta,\gamma\geq 0$. (If any of the these coefficients, say for example $\beta$, is negative, then, by making $c_{\phi^4}^{(2)}=0$ and $c_{\phi^4}^{(1)}=-c_{\phi^4}^{(3)}$ large, $\dot{c}_{B^2\phi^2 D^2}^{(1)}$ becomes negative.) The ellipses in this discussion indicate Wilson coefficients of other operator classes, which can be turned zero given that positivity can be saturated in the UV.

This result is what is shown in Tab.~\ref{tab:rges_rSMEFT}. We conclude that $\gamma_{c_{B^2\phi^2D^2}^{(1)},\, c_{\phi^4}^{(2)}}\geq \gamma_{c_{B^2\phi^2D^2}^{(1)},\, c_{\phi^4}^{(1)}}\geq \gamma_{c_{B^2\phi^2D^2}^{(1)},\, c_{\phi^4}^{(3)}}$; all them being non-negative.

In the same vein, from Eq.~\eqref{eq:pos4}, we infer that for the running of $B^2\phi^2 D^2$ by $e^2\phi^2 D^3$:
\begin{align}
 \dot{c}_{B^2\phi^2 D^2}^{(1)} &= -\delta (c_{e^2\phi^2 D^3}^{(1)}+c_{e^2\phi^2 D^3}^{(2)})+\cdots\,,
\end{align}
with $\delta\geq 0$. Thus, far from being arbitrary, these two anomalous dimensions must be equal as well as non-positive. From Eq.~\eqref{eq:pos4} itself, we can also derive that $\dot{c}_{e^2\phi^2 D^3}^{(1)}+\dot{c}_{e^2\phi^2 D^3}^{(2)}\geq 0$ for all $c_{\phi^4}^{(1)},c_{\phi^4}^{(2)},c_{\phi^4}^{(3)}$ satisfying Eq.~\eqref{eq:pos1} as well as for $c_{e^4 D^2}$ fulfilling Eq.~\eqref{eq:pos6}. This implies:
\begin{align}
 \dot{c}_{e^2\phi^2 D^3}^{(1)}+\dot{c}_{e^2\phi^2 D^3}^{(2)} &= (\epsilon+\zeta)c_{\phi^4}^{(1)} + (\epsilon+\zeta+\eta) c_{\phi^4}^{(2)} + \epsilon \,c_{\phi^4}^{(3)} - \theta \,c_{e^4 D^2}+\cdots\,,
\end{align}
with $\epsilon,\zeta,\eta,\theta\geq 0$.

Finally, from Eqs.~\eqref{eq:pos5} and \eqref{eq:pos4} and \eqref{eq:pos6} we derive:
\begin{align}
 \dot{c}_{e^2 B^2 D} = -\iota (c_{e^2\phi^2 D^3}^{(1)} + c_{e^2\phi^2 D^3}^{(2)}) -\kappa\, c_{e^4 D^2}+\cdots\,,
\end{align}
where $\iota,\kappa\geq 0$.

Following section~\ref{sec:dispersion_relations}, the crossed entries in Tab.~\ref{tab:rges_rSMEFT} can not be bounded on the basis of positivity, because $\gamma^\prime$ does not necessarily vanish.

\section{Extension to the full electroweak sector and non-renormalisation results}
\label{sec:full_smeft}
Let us now consider a more complete version of the SMEFT including $l$ and $W$, and with $g_2$ and $Y_e$ non-vanishing. (We consider still one single family and neglect colour together with quarks and gluons.) The number of SMEFT operators in this case rises to $\sim 200$. We avoid listing them all explicitly, but we use the notation and conventions of Ref.~\cite{Murphy:2020rsh}, from which the field content of operators is apparent; with the only exception that for the second $l^4 D^2$ interaction, we consider the more commonly used
\begin{equation}
\mathcal{O}_{l^4 D^2}^{(2)} = D_\nu(\overline{l}\gamma^\mu\sigma_I l)D^\nu (\overline{l}\gamma_\mu\sigma_I l)\,.
\end{equation}

On top of Eqs.~\eqref{eq:pos1}--\eqref{eq:pos6}, we obtain the following positivity bounds:

For $l_i\varphi_j\to l_i\varphi_j$, we get:
\begin{equation}
 -(c_{l^2\phi^2 D^3}^{(1)} + c_{l^2\phi^2 D^3}^{(2)} + c_{l^2\phi^2 D^3}^{(3)} + c_{l^2\phi^2 D^3}^{(4)})  \geq 0\,,\quad c_{l^2\phi^2 D^3}^{(3)} + c_{l^2\phi^2 D^3}^{(4)} - c_{l^2\phi^2 D^3}^{(1)} - c_{l^2\phi^2 D^3}^{(2)}  \geq 0\,.
\end{equation}

For $l_i B\to l_i B$, we get:
\begin{equation}
 -c_{l^2 B^2 D} \geq 0\,.
\end{equation}

For $e W^I\to e W^I$, we get:
\begin{equation}
 -c_{e^2 W^2 D} \geq 0\,.
\end{equation}

For $l_i W^I\to l_i W^I$, we get:
\begin{equation}
 -c_{l^2 W^2 D}^{(1)} \geq 0\,.
\end{equation}

For $l_i l_j\to l_i l_j$, we get:
\begin{equation}
 -c_{l^4 D^2}^{(2)}\geq 0\,,\quad -(c_{l^4 D^2}^{(1)}+c_{l^4 D^2}^{(2)})\geq 0\,.
\end{equation}

Finally, for $e l_i\to e l_j$, we obtain:
\begin{equation}
 -c_{l^2 e^2 D^2}^{(2)} \geq 0\,.
\end{equation}

We can now follow the same line of thought as in the previous section, and derive the correlations between different anomalous dimensions. Before proceeding this way, though, let us make an important remark.

The (non-trivial) vanishing entries in Tab.~\ref{tab:rges_rSMEFT} associated to the renormalisation of $\phi^4 D^4$ and $B^2\phi^2 D^2$ by $B\phi^4 D^2$ as well as of $B^2\phi^2 D^2$, $e^2\phi^2 D^3$ and $e^2B^2 D$ by $e^2 B\phi^2 D$, have been previously uncovered in the literature~\cite{Craig:2019wmo} on the basis of generalised unitarity~\cite{Bern:1994cg,Bern:1994zx} and on-shell amplitude methods~\cite{Cheung:2015aba}. Our result simply allows us to understand these zeros from a different perspective. However, we can further show that the correlations between different ADM entries can lead to new zeros in the appropriate basis of operators.

As a simple example, consider the renormalisation of $c_{B^2\phi^2 D^2}^{(1)}$ by $c_{e^2\phi^2 D^3}^{(1)}$ and $c_{e^2\phi^2 D^3}^{(2)}$. We can write it as:
\begin{equation}
 \dot{c}_{B^2\phi^2 D^2}^{(1)} = (-\delta\, -\delta)\cdot\vec{c}_{e^2\phi^2 D^3}^{\,\text{T}}\,,
\end{equation}
with $\vec{c}_{e^2\phi^2 D^3}=(c_{e^2\phi^2D^3}^{(1)}\, c_{e^2\phi^2 D^3}^{(2)})$.

Let us now consider a different basis for $e^2\phi^2 D^3$, consisting of two operators with Wilson coefficients $\tilde{c}_{e^2\phi^2 D^3}^{(1)}$ and $\tilde{c}_{e^2\phi^2 D^3}^{(2)}$, related to the previous ones by:
\begin{equation}\label{eq:1rot}
 \vec{c}_{e^2\phi^2 D^3} = P_{e^2\phi^2 D^3}\cdot \vec{\tilde{c}}_{e^2\phi^2D^3}\,,\quad P_{e^2\phi^2 D^3}= \left[\begin{smallmatrix} 1&0\\\\-1&1\end{smallmatrix}\right]\,.
\end{equation}
In this new basis, we have:
\begin{equation}
 \dot{c}_{B^2\phi^2 D^2}^{(1)} = (-\delta\, -\delta) \cdot P_{e^2\phi^2D^3}\cdot \vec{\tilde{c}}_{e^2\phi^2 D^3} = -\delta\, \tilde{c}_{e^2\phi^2 D^3}^{(2)} = (\fcolorbox{black}{LightCyan}{0}\, -\delta) \cdot\vec{\tilde{c}}_{e^2\phi^2 D^3} \,.
\end{equation}
We see that, in this new basis, there is a zero in the ADM. To the best of our knowledge, this sort of fixed point has not been described previously in the literature. Actually, the results of Refs.~\cite{Cheung:2015aba,Craig:2019wmo} prohibit the mixing between operators of certain weights, which are quantities that depend only on the number of particles and helicities of the corresponding operators. Hence, $e^2\phi^2 D^3$ mixing into $B^2\phi^2 D^2$ is \textit{a priori} always allowed.

The physical meaning of the transformation in Eq.~\eqref{eq:1rot} is the following. The \textit{two} operators in the class $e^2\phi^2 D^3$ are subject to a \textit{unique} positivity constraint; see Eq.~\eqref{eq:pos4}. Thus, we can search for a change of basis in which only one of the (new) Wilson coefficients appears in the positivity constraint. (In the case above, we have chosen $\tilde{c}_{e^2\phi^2 D^3}^{(2)}\leq 0$.) The remaining one can therefore have arbitrary sign and, consequently, its mixing into couplings whose running is bounded by positivity (for example $c_{B^2\phi^2 D^2}^{(1)}$) must vanish.

Reasoning alike for $l^2\phi^2 D^3$, we define:
\begin{equation}\label{eq:2rot}
 \vec{c}_{l^2\phi^2 D^3} = P_{l^2\phi^2 D^3}\cdot \vec{\tilde{c}}_{l^2\phi^2 D^3}\,,\quad P_{l^2\phi^2 D^3} = \bigg[\begin{smallmatrix} -2&1&-1&1 \\ 2&0&1&0 \\ 0&-1&-1&1\\0&0&1&0 \end{smallmatrix}\bigg]\,,
\end{equation}
with $\vec{c}_{l^2\phi^2 D^3}=(c_{l^2\phi^2 D^3}^{(1)}\,c_{l^2\phi^2 D^3}^{(2)}\,c_{l^2\phi^2 D^3}^{(3)}\,c_{l^2\phi^2 D^3}^{(4)})$ and similarly for the tilde counterpart. In this case, we have that $\tilde{c}_{l^2\phi^2 D^3}^{(2)}\leq 0$ and $\tilde{c}_{l^2\phi^2 D^3}^{(4)}\leq 0$, whilst the first and third ones are unconstrained.

In the new basis defined by Eqs.~\eqref{eq:1rot} and \eqref{eq:2rot}, the relevant part of our SMEFT ADM looks as in Tab.~\ref{tab:rges_with_zeros}.
\begin{table}[t]
 \begin{center}
  \resizebox{\textwidth}{!}{
  \begin{tabular}{ccccccccccccccc}
   \toprule 
   & \textbf{$c_{\phi^4 D^4}^{(1)}$} & \textbf{$c_{\phi^4 D^4}^{(2)}$} & \textbf{$c_{\phi^4 D^4}^{(3)}$} & \textbf{$\tilde{c}_{e^2\phi^2 D^3}^{(1)}$} & \textbf{$\tilde{c}_{e^2\phi^2 D^3}^{(2)}$} & \textbf{$\tilde{c}_{l^2\phi^2 D^3}^{(1)}$} & \textbf{$\tilde{c}_{l^2\phi^2 D^3}^{(2)}$} & \textbf{$\tilde{c}_{l^2\phi^2 D^3}^{(3)}$} & \textbf{$\tilde{c}_{l^2\phi^2 D^3}^{(4)}$} & \textbf{$c_{e^4 D^2}$} & \textbf{$c_{l^4 D^2}^{(1)}$} & \textbf{$c_{l^4 D^2}^{(2)}$} & \textbf{$c_{l^2 e^2 D^2}^{(1)}$} & \textbf{$c_{l^2 e^2 D^2}^{(2)}$} \\[0.5cm]
   \boldmath{$c_{B^2\phi^2 D^2}^{(1)}$} &  \textcolor{blue}{$\bm{+}$} &  \textcolor{blue}{$\bm{+}$} &  \textcolor{blue}{$\bm{+}$} & \fcolorbox{black}{LightCyan}{$\,0\,$} &  \textcolor{blue}{$\bm{-}$} & \fcolorbox{black}{LightCyan}{$\,0\,$} &  \textcolor{blue}{$\bm{-}$} & \fcolorbox{black}{LightCyan}{$\,0\,$} &  \textcolor{blue}{$\bm{-}$} & 0 & 0 & 0 & 0 & 0 \\ [0.2cm]
   \boldmath{$c_{W^2\phi^2 D^2}^{(1)}$} &  \textcolor{blue}{$\bm{+}$} &  \textcolor{blue}{$\bm{+}$} &  \textcolor{blue}{$\bm{+}$} & 0 & 0 & \fcolorbox{black}{LightCyan}{$\,0\,$} &  \textcolor{blue}{$\bm{-}$} & \fcolorbox{black}{LightCyan}{$\,0\,$} &  \textcolor{blue}{$\bm{-}$} &  0 & 0  & 0 & 0 & 0\\ [0.2cm]
   \boldmath{$\tilde{c}_{e^2\phi^2 D^3}^{(2)}$} &  \textcolor{blue}{$\bm{+}$} &  \textcolor{blue}{$\bm{+}$} &  \textcolor{blue}{$\bm{+}$} & \textcolor{gray}{$\times$} & \textcolor{gray}{$\times$} & \fcolorbox{black}{LightCyan}{$\,0\,$} &  \textcolor{blue}{$\bm{-}$} & \fcolorbox{black}{LightCyan}{$\,0\,$} &  \textcolor{blue}{$\bm{-}$} &  \textcolor{blue}{$\bm{-}$} & 0 & 0 & \fcolorbox{black}{LightCyan}{$\,0\,$} &  \textcolor{blue}{$\bm{-}$} \\ [0.2cm]
   \boldmath{$\tilde{c}_{l^2\phi^2 D^3}^{(2)}$} &  \textcolor{blue}{$\bm{+}$} &  \textcolor{blue}{$\bm{+}$} &  \textcolor{blue}{$\bm{+}$} & \fcolorbox{black}{LightCyan}{$\,0\,$} &  \textcolor{blue}{$\bm{-}$} & \textcolor{gray}{$\times$} & \textcolor{gray}{$\times$} & \textcolor{gray}{$\times$} & \textcolor{gray}{$\times$} & 0 &  \textcolor{blue}{$\bm{-}$} &  \textcolor{blue}{$\bm{-}$} & \fcolorbox{black}{LightCyan}{$\,0\,$} &  \textcolor{blue}{$\bm{-}$} \\ [0.2cm]
   \boldmath{$\tilde{c}_{l^2\phi^2 D^3}^{(4)}$} &  \textcolor{blue}{$\bm{+}$} &  \textcolor{blue}{$\bm{+}$} &  \textcolor{blue}{$\bm{+}$} & \fcolorbox{black}{LightCyan}{$\,0\,$} &  \textcolor{blue}{$\bm{-}$} & \textcolor{gray}{$\times$} & \textcolor{gray}{$\times$} & \textcolor{gray}{$\times$} & \textcolor{gray}{$\times$} & 0 &  \textcolor{blue}{$\bm{-}$} &  \textcolor{blue}{$\bm{-}$} & \fcolorbox{black}{LightCyan}{$\,0\,$} &  \textcolor{blue}{$\bm{-}$} \\ [0.2cm]
   \boldmath{$c_{e^2B^2 D}$} & 0 & 0 & 0 & \fcolorbox{black}{LightCyan}{$\,0\,$} &  \textcolor{blue}{$\bm{-}$} & 0 & 0 & 0 & 0 &  \textcolor{blue}{$\bm{-}$} & 0 & 0 & \fcolorbox{black}{LightCyan}{$\,0\,$} &  \textcolor{blue}{$\bm{-}$} \\ [0.2cm]
   \boldmath{$c_{l^2 B^2 D}$} & 0 & 0 & 0 & 0 & 0 & \fcolorbox{black}{LightCyan}{$\,0\,$} &  \textcolor{blue}{$\bm{-}$} & \fcolorbox{black}{LightCyan}{$\,0\,$} &  \textcolor{blue}{$\bm{-}$} & 0 &  \textcolor{blue}{$\bm{-}$} &  \textcolor{blue}{$\bm{-}$} & \fcolorbox{black}{LightCyan}{$\,0\,$} &  \textcolor{blue}{$\bm{-}$} \\ [0.2cm]
   \boldmath{$c_{e^2W^2 D}$} & 0 & 0 & 0 & \fcolorbox{black}{LightCyan}{$\,0\,$} &  \textcolor{blue}{$\bm{-}$} & 0 & 0 & 0 & 0 & 0 & 0 & 0 & \fcolorbox{black}{LightCyan}{$\,0\,$} &  \textcolor{blue}{$\bm{-}$} \\ [0.2cm]
   \boldmath{$c_{l^2W^2 D}^{(1)}$} & 0 & 0 & 0 & 0 & 0 & \fcolorbox{black}{LightCyan}{$\,0\,$} &  \textcolor{blue}{$\bm{-}$} & \fcolorbox{black}{LightCyan}{$\,0\,$} &  \textcolor{blue}{$\bm{-}$} & 0 &  \textcolor{blue}{$\bm{-}$} &  \textcolor{blue}{$\bm{-}$} & 0 & 0 \\ [0.2cm]
   \boldmath{$c_{l^2 e^2 D^2}^{(2)}$} & 0 & 0 & 0 & \fcolorbox{black}{LightCyan}{$\,0\,$} &  \textcolor{blue}{$\bm{-}$} & \fcolorbox{black}{LightCyan}{$\,0\,$} &  \textcolor{blue}{$\bm{-}$} & \fcolorbox{black}{LightCyan}{$\,0\,$} &  \textcolor{blue}{$\bm{-}$} &  \textcolor{blue}{$\bm{-}$} &  \textcolor{blue}{$\bm{-}$} &  \textcolor{blue}{$\bm{-}$} & \textcolor{gray}{$\times$} & \textcolor{gray}{$\times$} \\ [0.2cm]
   \bottomrule
  \end{tabular}
  }
 \end{center}
 \caption{\it Structure of the SMEFT ADM in the subspace of operators constrained by positivity. The $\textcolor{blue}{\bm{+}}$ $(\textcolor{blue}{\bm{-}})$ implies that the corresponding entry is $\geq 0$ ($\leq 0$). The non-shaded zeros are trivial; see the text for details.}\label{tab:rges_with_zeros}
\end{table}
In this occasion, we have simply specified the signs and zeros, but we should keep in mind that some more relevant information (as for example the respective size of certain anomalous dimensions) can be also unraveled by this analysis. Note in addition that the $\textcolor{blue}{\bm{+}}$ and $\textcolor{blue}{\bm{-}}$ entries indicate definite sign irrespective of the actual values of the SM couplings, meaning that they could be proportional to combinations like for example $g_2^2+Y_e^2$, but not $g_2^2-Y_e^2$.  

\section{Conclusions}
\label{sec:conclusions}
We have derived a number of restrictions on the anomalous dimensions of the SMEFT at dimension eight, relying uniquely on the crossing symmetry, analyticity and positivity of the imaginary part of two-to-two scattering amplitudes in the forward limit. In short, our results are based on the following findings.

\textit{(i)} The Wilson coefficients $c_i$ of a number of dimension-eight operators of the form $\mathcal{O}_i=c_i\psi^2\psi^{\prime 2}$ ($\psi$ represents a generic light field, either fermion or boson) generated at tree level in well-behaved UV completions of the SMEFT, are subject to positivity constraints of the sort $c_i>0$. \textit{(ii)} The running of any such dimension-eight operator $\mathcal{O}_i$ as triggered by any other tree-level dimension-eight interaction $\mathcal{O}_j$ fulfills $\gamma_{ij}c_j\leq 0$ whenever the renormalised operator $\mathcal{O}_i$ involves at least one field not contained in $\mathcal{O}_j$ (for example, $e^2 B^2 D$ renormalised by $e^2\phi^2 D^3$; but not $e^4 D^2$ renormalised by $e^2\phi^2 D^3$). \textit{(iii)} If $c_j$ is itself bounded by positivity, $c_j\geq 0$, then $\gamma_{ij}\leq 0$; otherwise, namely if $c_j$ can have either sign, then necessarily $\gamma_{ij}=0$.
This way, restricting to the electroweak sector of the SMEFT with only one flavor, and in the appropriate basis of operators,  we have found 52 elements of the ADM that must have definite sign (either non-positive or non-negative), as well as 24 non-trivial zeros. Moreover, despite not being emphasised as much, we have found inequalities involving the aforementioned anomalous dimensions themselves.

We can envisage different future directions. To start with, it would be desirable to cross-check our results by explicit calculation. Also, we can envision applying these findings to phenomenological studies where the running of dimension-eight operators might be important~\cite{Panico:2018hal,Ardu:2021koz,Corbett:2021eux,Alioli:2022fng,Asteriadis:2022ras}. Likewise, it would be interesting to extend these results to the full SMEFT (that means, including colour and flavour) as well as to the LEFT~\cite{Jenkins:2017jig} and other EFTs, as for example those involving sterile neutrinos~\cite{delAguila:2008ir,Chala:2020vqp} or axion-like particles~\cite{Gripaios:2016xuo,Chala:2020wvs}, with the aim of understanding better the quantum structure of these theories. Finally, it might be worth exploring whether our results hold also for mixing induced by loop-operators like $B^2\phi^2 D^2$, $B^4$ or $e^2 B^2 D$. As a matter of fact, there exist tree-level dimension-five UV completions of these~\cite{Bi:2019phv}, which are perturbatively unitary and for which the Froissart bound is also satisfied.

\section*{Acknowledgments}
I am grateful to Renato Fonseca for sharing a basis of redundant operators of the rSMEFT needed for the computations in Appendix~\ref{app:xcheck}, obtained within a (not-yet-published) update of \texttt{Sym2Int}~\cite{Fonseca:2019yya}. I am also thankful
to Mario Herrero-Valea, Guilherme Guedes, Maria Ramos and Jose Santiago for useful discussions. I would also like to thank the organisers and participants of SMEFT-Tools 2022 for fostering discussions valuable for this work. V2: I am thankful to Xu Li and Jiayin Gu for spotting missing terms in the derivation of Eq.~\ref{eq:positive_beta}. This work is supported by SRA under grants 
PID2019-106087GB-C21 and PID2021-128396NB-I00, by the Junta de Andaluc\'ia grants FQM 101, A-FQM-211-UGR18, P21-00199 and P18-FR-4314 (FEDER), as well as by the Spanish MINECO under the Ram\'on y Cajal programme.
%
%
\appendix
\section{Explicit computation in the reduced SMEFT}
\label{app:xcheck}
In this section, we provide the explicit result for the blue entries of the ADM of Tab.~\ref{tab:rges_rSMEFT}, but including also contributions from the Yukawa $Y_e$. Some of these were computed already in Ref.~\cite{DasBakshi:2022mwk}. For the rest, we proceed as in that reference, namely computing all relevant one-particle-irreducible (1PI) diagrams off-shell, extracting the divergences and projecting them onto a Green's basis of operators~\cite{Chala:2021cgt}. This process is tedious and largely prone to error, thus we rely on \texttt{FeynArts}~\cite{Hahn:2000kx} and \texttt{FormCalc}~\cite{Hahn:1998yk}, with partial cross-checks from \texttt{matchmakereft}~\cite{Carmona:2021xtq} as well.

\begin{figure}[t]
 \begin{center}
 \includegraphics[width=0.23\columnwidth]{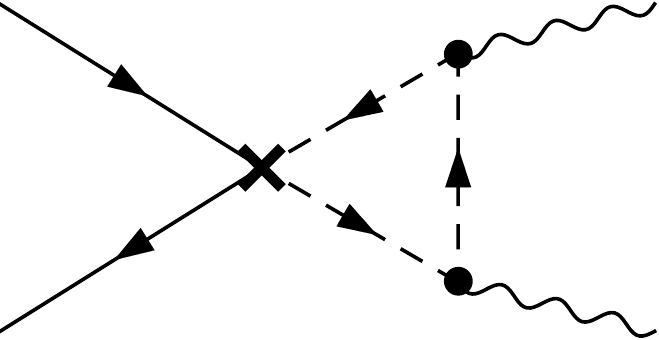}
 \hspace{0.1cm}
 \includegraphics[width=0.23\columnwidth]{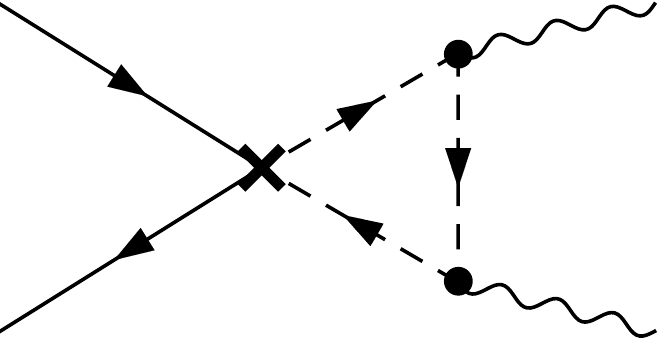}
 \hspace{0.1cm}
 \includegraphics[width=0.23\columnwidth]{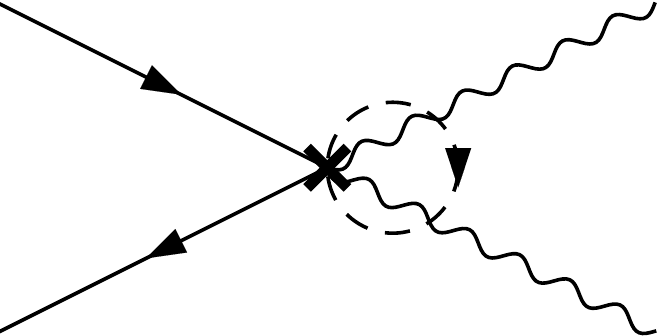}
 \hspace{0.1cm}
 \includegraphics[width=0.23\columnwidth]{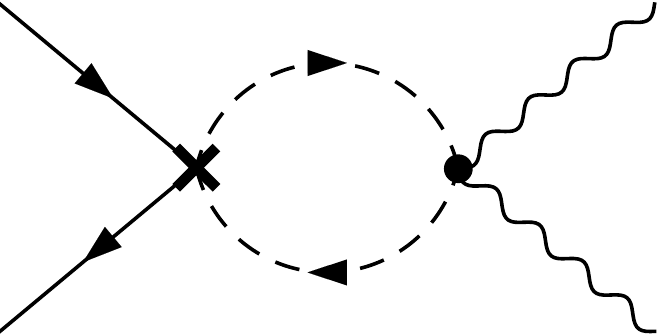}\\[0.3cm]
 \includegraphics[width=0.23\columnwidth]{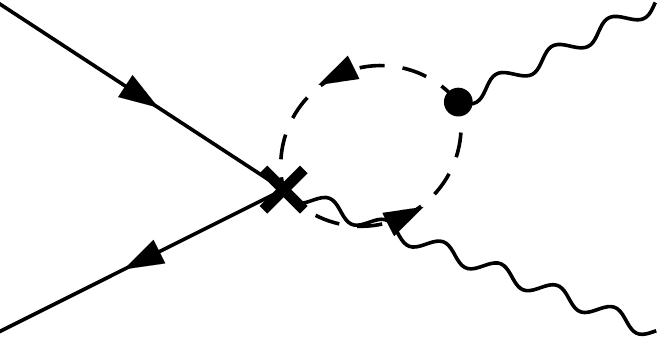}
 \hspace{0.1cm}
 \includegraphics[width=0.23\columnwidth]{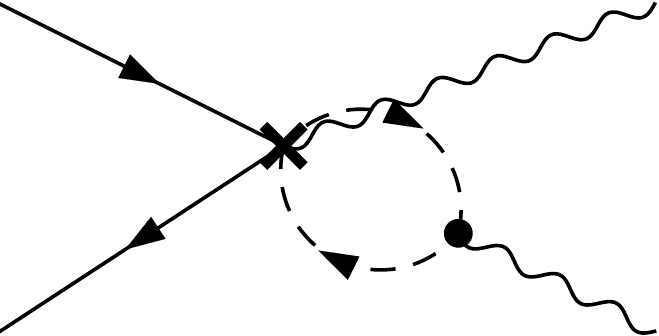}
 \hspace{0.1cm}
 \includegraphics[width=0.23\columnwidth]{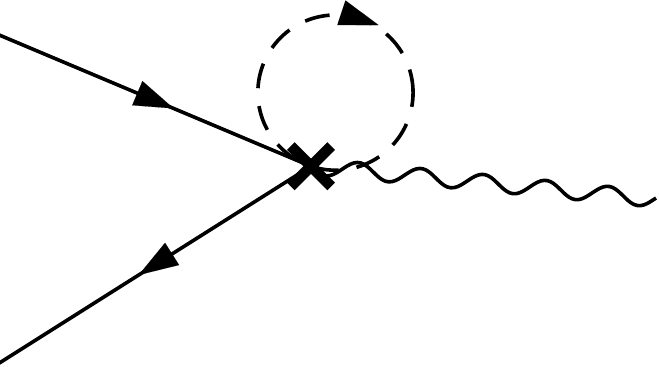}
 \hspace{0.1cm}
 \includegraphics[width=0.23\columnwidth]{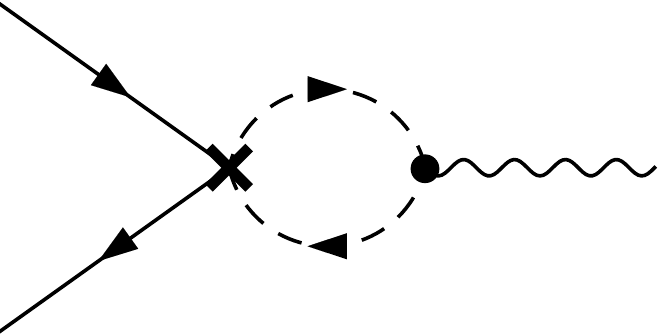}
 \end{center}
 \caption{\it 1PI diagrams relevant for the off-shell renormalisation of $e^2 B^2 D$ operators (first six diagrams) and of redundant $e^2 B D^3$ interactions (last two diagrams).}\label{fig:diagrams_xcheck}
\end{figure}

We work in dimensional regularisation with space-time dimension $D=4-2\varepsilon$. 
As a matter of example, let us detail the computation of the one-loop running of $e^2 B^2 D$ by $e^2\phi^2 D^3$. Because we work off-shell, loops of $e^2\phi^2 D^3$ can generate divergences for both physical and redundant $e^2 B^2 D$ terms as well as for (redundant) $e^2 B D^3$ interactions; see Fig.~\ref{fig:diagrams_xcheck}. Explicitly, the relevant divergent Lagrangian reads:
\begin{align}
 \mathcal{L}_\text{div} = \frac{g_1}{384\pi^2\varepsilon} &\bigg[4 g_1 (c_{e^2\phi^2 D^3}^{(1)}+c_{e^2\phi^2 D^3}^{(2)})\mathcal{O}_{e^2 B^2 D}-2g_1\,(c_{e^2\phi^2 D^3}^{(1)}+c_{e^2\phi^2 D^3}^{(2)})\mathcal{O}_{e^2 B^2 D}^{\prime}\\\nonumber
 &+2 (c_{e^2\phi^2 D^3}^{(1)}-c_{e^2\phi^2 D^3}^{(2)})\mathcal{O}_{e^2 BD^3}\bigg]\,,
\end{align}
where
\begin{align}
 \mathcal{O}_{e^2 B^2 D}^{\prime} &= \ii (\overline{e}\slashed{D} e) B_{\mu\nu} B^{\mu\nu}+\text{h.c.}
\end{align}
and
\begin{align}
 \mathcal{O}_{e^2 B D^3} &= (\overline{e}\gamma^\nu e) D^2 D^\mu B_{\mu\nu}\,.
\end{align}

Other operators are either non-renormalised, or irrelevant for the anomalous dimension under consideration. 

By using the rSMEFT equations of motion, the first redundant operator above enters the class $le\phi B^2$ (because $\ii\slashed{D}e=Y_e^*\phi^\dagger l$), while the second one moves to the class $e^2 \phi^2 D^3$.
Thus, the divergence of $c_{e^2 B^2 D}$ in the physical basis is simply:
\begin{equation}
 \text{div}\,(c_{e^2 B^2 D}) = \frac{g_1^2}{96\pi^2} (c_{e^2\phi^2 D^3}^{(1)}+c_{e\phi^2 D^3}^{(2)})\,.
\end{equation}
From here, we get that:
\begin{equation}
 \dot{c}_{e^2 B^2 D} = -16\pi^2 \sum_{i} \alpha_i\, n_i \frac{\partial\, \text{div} (e^2 B^2 D)}{\partial \alpha_i} = -\frac{g_1^2}{3}(c_{e^2\phi^2 D^3}^{(1)}+c_{e^2\phi^2 D^3}^{(2)})\,.
\end{equation}
The $i$ in the sum runs over all necessary couplings, $\alpha_1,\alpha_2,\alpha_3=g_1,c_{e^2\phi^2 D^3}^{(1)},c_{e^2\phi^2 D^3}^{(2)}$; whereas $n_i$ stand for their corresponding classical anomalous dimensions: $1,2$ and $2$, respectively.
This result matches Tab.~\ref{tab:rges_rSMEFT} for $\kappa =g_1^2/3\geq 0$.
Proceeding this way for the remaining anomalous dimensions, we obtain:
\begin{align}
 \alpha = \frac{g_1^2}{6}\,,\quad \alpha+\beta = \frac{g_1^2}{3}\,,\quad \alpha+\beta+\gamma = \frac{g_1^2}{2}\,,\quad \delta = \frac{16 g_1^2}{3}\,;\quad\quad\nonumber\\[0.2cm]
 \epsilon = \frac{|Y_e|^2}{3}\,,\quad \epsilon+\zeta = \frac{2 |Y_e|^2}{3}\,,\quad \epsilon+\zeta+\eta = |Y_e|^2\,,\quad \theta = Y_e^2\,,\quad \iota = \frac{g_1^2}{3}\,.
\end{align}

Finally, in order to highlight how special the anomalous dimensions singled out in this work are, let us simply state that the sign of most of these quantities is in general ill-defined. For example, we have that
\begin{equation}
 \dot{c}_{e^2\phi^2 D^3} \propto (g_1^2-4 |Y_e|^2) c_{e^4 D^2}\,,
\end{equation}
wich can be either positive or negative depending on the value of $g_1/|Y_e|$. This also strengthens the idea, already mentioned in the conclusions, that our results provide, indirectly, information about the functional form the combination of SM gauge couplings involved in the restricted anomalous dimensions.

\bibliographystyle{style} 

\bibliography{refs} 

\end{document}